\documentclass[a4paper]{article}
\usepackage{Odyssey2018}
\usepackage{epsfig,amssymb,amsmath}
\usepackage[titletoc]{appendix}

\usepackage{makecell} 
\usepackage{multirow}
\usepackage{xcolor, colortbl}

\ninept

\title{A Study On Convolutional Neural Network Based End-To-End Replay Anti-Spoofing}

\makeatletter
\def\name#1{\gdef\@name{#1\\}}
\makeatother
\name{{\em Bhusan Chettri, Saumitra Mishra, Bob L. Sturm and Emmanouil Benetos}}

\address{School of Electronic Engineering and Computer Science  \\
Queen Mary University of London, United Kingdom \\
{\small \tt \{b.chettri,saumitra.mishra,b.sturm,emmanouil.benetos\}@qmul.ac.uk}}
\begin{document}
\maketitle

\begin{abstract}

The second Automatic Speaker Verification Spoofing and Countermeasures challenge (ASVspoof 2017) focused on ``replay attack'' detection. The best deep-learning systems to compete in ASVspoof 2017 used Convolutional Neural Networks (CNNs) as a feature extractor. In this paper, we study their performance in an end-to-end setting. We find that these architectures show poor generalization in the evaluation dataset, but find a compact architecture that shows good generalization on the development data. We demonstrate that for this dataset it is not easy to obtain a similar level of generalization on both the development and evaluation data. This leads to a variety of open questions about what the differences are in the data; why these are more evident in an end-to-end setting; and how these issues can be overcome by increasing the training data.

\end{abstract}

\section{Introduction}

The Automatic Speaker Verification Spoofing and Countermeasures Challenge (ASVspoof) focuses on techniques to make the automatic speaker verification (ASV) systems robust against \textit{spoofing}. Spoofing attack `fools' an ASV system by using a speech utterance that imitates the vocal characteristics of the real speaker.  The four commonly used methods to generate spoofed speech are (1) text-to-speech; (2) voice conversion; (3) mimicry; and (4) replay. 

The second version of the challenge, referred as ASVspoof 2017\footnote{organised at Interspeech 2017}, focused on text-dependent replay attack detection \cite{ASV2017plan,tomiSummaryPaper}. The replay attack is a simple spoofing method that involves recording the original speech (through a recording device, e.g., mobile phone) and then replaying it (through a playback device, e.g., speakers) to the biometric system. The challenge involved building an anti-spoofing system to identify an input speech utterance as genuine or spoofed. Building such a system is a challenging task. One reason could be the high quality of the replayed speech (if recorded and replayed through a high-quality device). Moreover, the ASVspoof 2017 challenge was particularly difficult as the challenge datasets were imbalanced (biased towards the `spoofed' class) and had a large number of mutually exclusive spoofing configurations.

One way to design a replay attack detection system is by hand-crafting features that capture the cues to differentiate between a genuine and a replayed signal. But, this feature extraction approach often requires domain expertise and is especially challenging for modelling high-dimensional data (e.g., images, audio). In another direction, researchers propose to train deep neural network models (DNNs) to learn the desired features automatically from data \cite{LeCun2015}. Recent results claim that with a `large' amount of training data\footnote{e.g., millions of images, several hundred hours of audio data} and `enough' computing resources\footnote{powerful GPU's}, the deep models out-perform the shallow models.

Many competing systems \cite{lavrentyeva2017audio, pindrop, resnet, replay_using_highFreqFeats,iiit_hyderabad} in the ASVspoof 2017 used DNN models. For example, the best systems in the challenge \cite{lavrentyeva2017audio} and \cite{pindrop} used deep Convolutional Neural Network (CNN) models to learn better feature representations. Later, the authors trained shallow classifiers (Gaussian Mixture Model (GMM) and Support Vector Machine (SVM), respectively) over the extracted features to discriminate between genuine and spoofed recordings. The best system from \cite{lavrentyeva2017audio}, that is a fusion of three models (two of which use DNNs), achieves a remarkable performance (EER = 6.73) on the evaluation dataset. The success of deep learning for the ASVspoof 2017 inspires our research to analyse and understand the behaviour of these models \cite{Montavon2018}. For example, given a trained neural network model, we plan to visualise the features that influence its decisions. A precursor to analyse these models is to train one that performs `fairly' (better than the baseline) on the evaluation dataset. 

In this work we report our experiments and challenges to design a deep anti-spoofing system that is trained and evaluated on ASVspoof database. We first describe our work to replicate the state-of-the-art system \cite{lavrentyeva2017audio} in an end-to-end setting. We train an end-to-end network as it is a preliminary requirement to use the feature visualisation and model analysis techniques from the literature \cite{Simonyan2014,Zeiler2014,Mishra2017}. We found that our CNN-based model generalises in the development dataset, but consistently underperforms in the evaluation dataset (section 3). We later explain our experiments to find a suitable architecture that generalises well to the unseen data (section 4). We explored a number of architectures, including the second best system in the challenge \cite{pindrop}. But, the performance on the evaluation dataset is always poor (EER $> 26\%$). This raises several interesting questions about the possible differences in the dataset and why are they more evident in an end-to-end setting and what are the possible ways to tackle this problem. We also propose a novel CNN architecture for the spoofing detection task that has about $5k$ free parameters.

\section{Background} \label{background}
In this section we briefly introduce ASVspoof2017 challenge, the spoofing dataset and some of the top performing deep-learning systems.

\subsection{The ASVspoof 2017 Challenge and the Database}

\begin{table}[t!]   
	\caption{\it The ASVspoof 2017 database statistics.}
    \vspace{1.5mm}
	\centering	
	\scalebox{1}{    
		\begin{tabular}{|c|c|c|c|c|}
			\hline			
            subset  & \# spkrs & \# genuine &\# spoofed  &dur (hr)  \\
			\hline \hline
			train	& 10 & 1508  & 1508 &2.22  \\ 
			dev	& 8	 & 760	& 950  &1.44   \\ 
			eval	& 24 & 1298	&12922 &11.95  \\
			\hline    
	\end{tabular}}
	\label{tab:dataset1}
\end{table}

The ASVspoof 2017 challenge was held as a special session at the Interspeech 2017 conference. The challenge witnessed a huge participation with a total of 49 submitted systems. The first ASV spoofing challenge was held in 2015 that focused on text independent text-to-speech and voice conversion spoofing. The ASVspoof 2017 focused on text-dependent replay attack detection `in the wild' with varying acoustic conditions \cite{ASV2017plan}. Given a recorded speech utterance $s$, the main goal of the ASVspoof 2017 challenge is to build an anti-spoofing system that determines if $s$ is a genuine speech.

In Table \ref{tab:dataset1} we show the ASVspoof 2017 database statistics. More details on the database can be found in \cite{kinnunen2017reddots}. Our prior work in \cite{bhusanIcassp} reports issues found in this database. The model predictions are highly influenced by the initial silence frames of zeros present in the genuine signals but missing in the spoofed counterpart. We also found that the two audio files $T\_1001658.wav$ and $T\_1000150.wav$ do not contain any speech recording. Therefore, we have removed them in our study. Recently, an updated ASVspoof 2017 database version 2 has been released online\footnote{https://datashare.is.ed.ac.uk/handle/10283/3017}. Our work in this paper, however, is based on the version 1 database.

In Table 2 we provide insights on how spoofed audio files are distributed between the training and the development set. A total of fifteen playback devices (P01-P15), sixteen recording devices (R01-R016) and six environments (E01-E06) are used to develop the ASVspoof 2017 replay database \cite{tomiSummaryPaper}. On the development set we find two new environments (E01 and E03); five new playback devices and six recording devices that do not appear in the training set. We find three spoofing configurations\footnote{Spoofing configuration refers to a unique combination of playback, recording device and the environment where the audio is replayed} in the training set and nine in the development set. A spoofing configuration `E02 P02 R04' seems to appear in both the training and the development set. 

\begin{table}[t!]	
\caption{\it Spoofing configuration statistics on the ASVspoof 2017 version 1 database. 
The number inside the bracket indicates the number of audio files. The letters t,d,e refers to training, development and evaluation subset. env, pd and rd denotes spoofing environment, playback and recording device respectively.}    
	\centering
	\vspace{1.5mm}
	\scalebox{0.87}{
		\begin{tabular}{|l|l|l|l|l|}
			\hline			
            sub  & \# env & \# pd &\# rd  &\# configurations \\ 
			\hline \hline
			t	&E02(335)  &P02(335) &R04(1508)  &E02 P02 R04 (335)  \\ 
				&E05(1173)  & P05(1169)  &           & E05 P05 R04 (1168)  \\ 
     			&           & P10(4)     &   & E05 P10 R04 (4)  \\ [1ex]  
     		\hline	
			d	& E01(95)	 &P02(95) 	&R02(95)    &E02 P02 R04 (95)  \\ 
							& E02(190)	 &P04(95) 	&R04(95)  & E05 P08 R07 (95) \\ 
			             	& E03(285)	 &P07(95) 	&R06(95)  & E05 P08 R11 (95) \\ 
			            	& E05(380)	 &P09(95) 	&R03(95)  & E03 P15 R08 (95) \\ 
				            & 	         &P15(285) 	&R07(190)  & E03 P15 R07 (95)  \\ 
				            & 	         &P08(285) 	&R08(190)  & E03 P15 R11 (95)  \\ 
				            & 	         & 	        &R11(190)  & E05 P04 R03 (95) \\ 
				            & 	         & 	        &          & E02 P07 R02 (95)  \\
				            & 	         & 	        &  & E05 P08 R08 (95) \\ 
							& 	         & 	        &  & E01 P09 R06 (95) \\ 				            						
			\hline    			
            e	&\multicolumn{3}{c}{meta-data not available} &   \\ 
			\hline    
	\end{tabular}}
	\label{tab:dataset2}
\end{table}

\subsection{Published Deep-Learning Systems}

We now provide a short description of the published deep-learning systems on the ASVspoof 2017 database. 


\begin{itemize}
 \item A \cite{lavrentyeva2017audio}: This system used score-level fusion of three systems. The first is a GMM trained on features extracted from a CNN. The second is an i-vector based SVM system trained on linear prediction cepstral coefficients and the third system is an end-to-end CNN-RNN system.
 
 \item B \cite{pindrop}: They train a CNN to model spoofing configuration on tandem features obtained by combining Constant Q Cepstral Coefficient (CQCC) with High Frequency Cepstral Coefficient (HFCC). Then they use it as a feature extractor to obtain high dimensional features on which a binary SVM classifier is trained to discriminate genuine and a spoofed class.

 \item C \cite{resnet}: This system employed score-level fusion of three systems. The first is a GMM trained on the CQCC features. The second and third systems are residual neural networks (ResNet) trained on the MFCC and CQCC features respectively.
  
 \item D \cite{resnet_dataAugmentation}: This system used score-level fusion of three systems. The first is a GMM system trained on CQCC features. The second is also a GMM system but trained on CQCC features obtained from the augmented data. The third system is a residual neural network.

\item E \cite{iiit_hyderabad}: This system use score-level fusion of GMM and Bi-directional long short term memory network (BLSTM). They use their proposed Single Frequency Filter Cepstral Coefficients (SFFCC) based delta-features to train the GMM and BLSTM models.

 \item  $LCNN_{FFT}$ \cite{lavrentyeva2017audio}: This is one of the sub-system of A \cite{lavrentyeva2017audio}, that use features extracted from a CNN to train a single-component GMM to model spoofed and genuine classes. This system has outperformed all other systems on the evaluation data by a large margin.
  
\end{itemize}

\begin{table}
\caption{\it Performance (EER \%) of the published deep-learning systems on the development and evaluation data.}
\centering
\vspace{1.5mm}
  \begin{tabular}{|c|c|c|}
  \hline  
  system &dev & eval \\    
  \hline  \hline
  A \cite{lavrentyeva2017audio}             & 3.95 & 6.73    \\
  B \cite{pindrop}                          & 7.6  & 11.5     \\
  C \cite{resnet}                           & 2.58 & 13.29   \\
  D \cite{resnet_dataAugmentation}          & 3.52 & 16.39   \\  
  E \cite{iiit_hyderabad}                   & 2.21 & 17.82   \\ [1ex]
  $LCNN_{FFT}$ \cite{lavrentyeva2017audio}  & 4.53 & 7.34   \\ 
  \hline
  \end{tabular}
  \label{deep_system_results}
\end{table}

\begin{figure*}[t!] 
  \centering    
  \includegraphics[width=\linewidth]{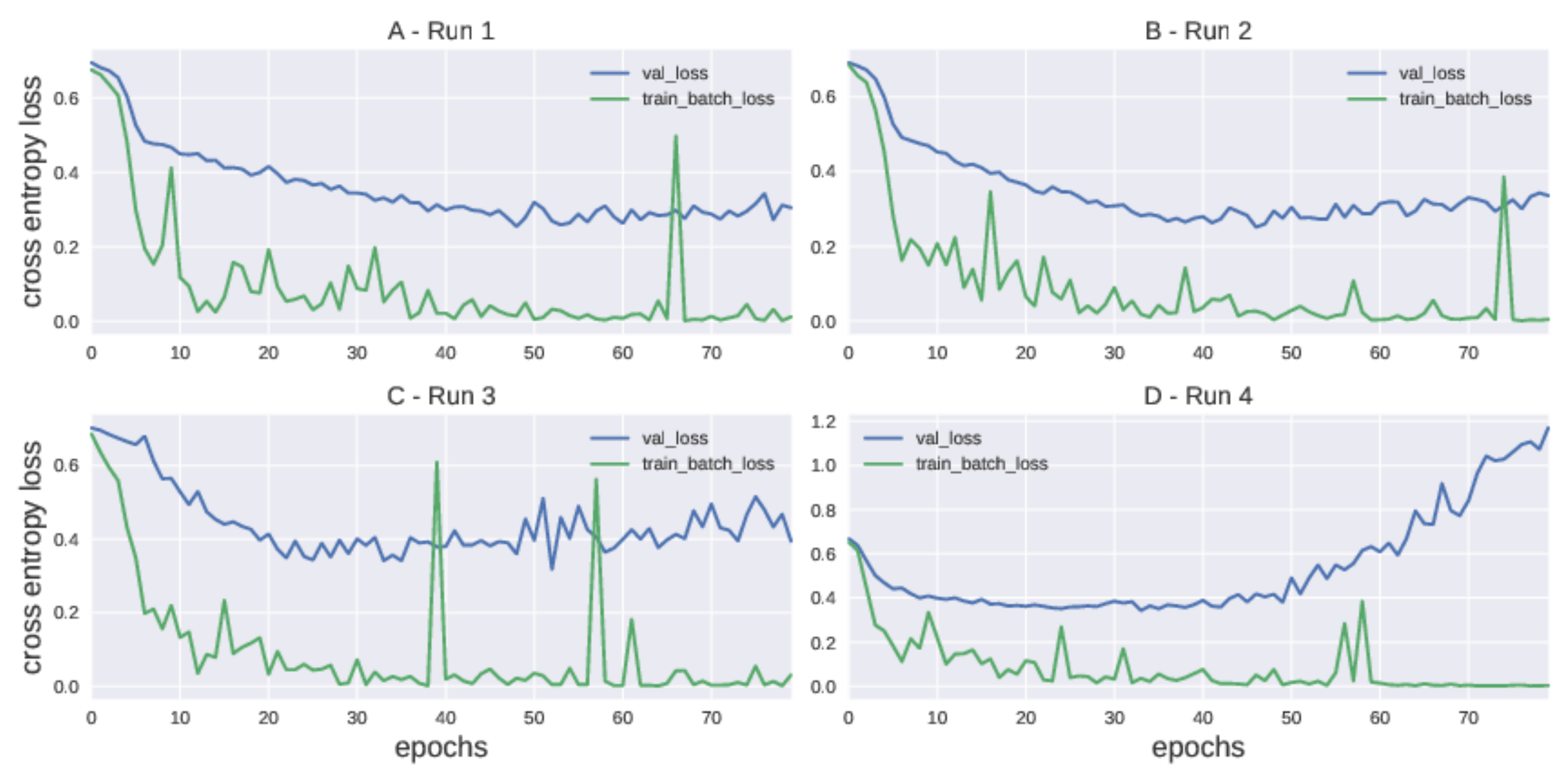}
  
  \caption{\it Cross entropy loss on the training and development data for four different runs of replicating the $LCNN_{FFT}$ of the state-of-the-art \cite{lavrentyeva2017audio}. A dropout of 70\% is applied during training. We observe some consistency on the loss pattern on the development data for different runs except the run 4.}
  \label{loss_curves}
\end{figure*}

In Table \ref{deep_system_results}, we present the results of these systems on the development and evaluation data. We observe a remarkable performance by system A, the state-of-the-art \cite{lavrentyeva2017audio}. The top two systems (A and B) show similar-level of generalization \footnote{the gap between the EER on the development and the evaluation dataset is nearly the same}on the development and the evaluation test sets which, however, contradicts with performance shown by the other systems C, D and E.



\subsection{Discussion and Motivation} \label{questions_for_study}

Usually the success of deep-learning systems is attributed to the availability of large training data. However, within the context of ASVspoof 2017, where the training data is significantly less than the test data, training a deep neural network to be able to achieve good generalization can be challenging. Therefore, we outline following questions that motivates our research work.

\begin{enumerate}    
	\item Using only the available training and development data, is it possible to train an end-to-end CNN that generalizes well to the evaluation dataset (EER = $7\%$ (approx.))?	
    
    \item  Why is there inconsistency in model generalization between the development and the evaluation datasets? Would this appear in end-to-end CNN systems too?
	
	\item Lastly, given such a small training data, can we design a deep architecture with as fewer trainable parameters that neither underfits nor overfits on training data?
    
\end{enumerate}


This work tries to seek answers to the above questions and discusses the possible outcome. For this, we start with replicating the CNN architecture of the state-of-the-art.


\section{Replicating the state-of-the-art CNN \cite{lavrentyeva2017audio}} \label{replicating}

We describe our experiments in replicating the best CNN system, $LCNN_{FFT}$, of the state-of-the-art \cite{lavrentyeva2017audio}. We train our CNN using their network parameterization but, with different input representation. We use 2048 FFT points and 2048 window size with a hop of 10ms to compute the spectrograms. We use Librosa\footnote{http://librosa.github.io} library for computing the spectrograms. Therefore, our input spectrogram is $ 400 \times 1025$ (time $\times$ frequency) dimension in comparison to $400 \times 864$ used by $LCNN_{FFT}$ \cite{lavrentyeva2017audio}. We envisage that use of 2048 as FFT size should not deteriorate performance dramatically. The details of the $LCNN_{FFT}$ architecture can be found in \cite{lavrentyeva2017audio}.

\subsection{Model Training and Testing} \label{cnn_training_style0}

The input to the network is a mean-variance normalized log power magnitude spectrogram of $400 \times 1025$ (time $\times$ frequency) dimension, where time denotes number of frames and frequency the number of bins. We compute mean and variance on the training data. We initialize our network weights using Xavier initialization \cite{glorot} and bias with zero. The network is trained to optimize the cross entropy loss between a genuine and a spoofed class. As specified in \cite{lavrentyeva2017audio}, we use max-feature-map (MFM) non-linearity, learning rate of 1e-4, batch size of 32, 0.9 momentum with ADAM optimizer. However, the default parameter value of epsilon did not work and we used 0.1 for epsilon. A dropout of 70\% to the inputs of the first fully connected layer is used during model training. We use tensorflow \cite{tensorflow2015-whitepaper} framework for implementation. We used early stopping as a terminating criterion: if the validation loss do not improve for 30 epochs then we abort the training loop. We use a maximum of 300 training epochs and chose the model that show the best performance on the validation data. 

At inference time, for each audio spectrogram the model outputs a posterior probability distribution for genuine and spoofed classes. We convert the posterior probability into log likelihoods ratio and compute the EER using the official Bosaris toolkit \cite{bosaris2013brummer}.

\subsection{Results and Discussion}

\begin{table}
 \caption{\it Performance (EER\%) of our replicated CNNs under two settings: end-to-end and using 1-mixture GMM trained on CNN features. * indicates EER on the training data. System F is trained on the development data and all other systems are trained on the training data.}  
 \centering
    \vspace{1.5mm}
	\scalebox{1}{
		\begin{tabular}{|c|c|c|c|c|}
			\hline
			\multirow{ 2}{*}{System} & \multicolumn{2}{c|}{End-to-End} & \multicolumn{2}{c|}{GMM}  \\	                   &dev & eval & dev &eval \\ [1ex]
            \hline \hline
			A & 9.04 &32.02 & 9.49  & 34  \\
            B & 9.30 &37.67 & 10.46 & 39  \\
            C & 8.01 &30.96 & 9.4  & 34.24 \\
            D & 14.11 &36.97 & 15.6 & 36.9\\ 			                                    	
            E & 9.11  &37.34 & 10.78 &35.66 \\ [1ex] 			
            \hline            
			F & 2.17* &38.83  &\multicolumn{2}{c|}{na}  \\
			\hline						
	\end{tabular}}
	\label{state_of_art}
\end{table}

\begin{table}[t!]
\caption{\it Performance (EER\%) of our best end-to-end CNN systems on the development and evaluation data, ASVspoof 2017 database version 1. * denotes the best replicated system using $LCNN_{FFT}$.}
\vspace{1.5mm}
\centering
  
  \begin{tabular}{|c|c|c|c|c|}
  \hline
  System & dev & eval& \# params & \makecell{\# params \\ after dropout}   \\ [1ex]
  \hline \hline
  C * & 8.01   & 30.96 & 371K  & 138K   \\
  Model 1 & 5.47  & 25.28 & 4M  & 600K  \\
  Model 2 & 4.52  & 34.91 & 68K & 35K \\
  Model 3 & 4.98  & 33.11 & 7682 & 5089 \\    
  \hline  
  \end{tabular}
  \label{all_results}
\end{table}

In Figure \ref{loss_curves} we show the loss curves of our four systems (A-D), depicting four different runs of model training. All these systems are trained on the training data and validated on the development data. For comparison with the state-of-the-art, we also trained a one-component GMM on the 32-dimensional features extracted from our trained CNNs. We present these results in Table \ref{state_of_art}. None of our systems could achieve a performance closer to what $LCNN_{FFT}$ \cite{lavrentyeva2017audio} reported on the evaluation data. Our best performing system C show an EER of 8.01\% and 30.96\% on the development and evaluation data under end-to-end condition and 9.4\% and 34.24\% using GMMs. The performance shown by our end-to-end models and the GMMs do not show large difference. We further see an interesting observation when we trained our CNN, system F, on the development data and used training data for model validation. The model show an impressive generalization on the training data with an EER of about 2\% but give a worse EER of 39\% on the evaluation data.

Therefore these experiments suggest that it is quite difficult to achieve the same-level of generalization between the two test sets. In the next section we investigate new CNN architecture to see if we can achieve same-level of generalization on both the development and the evaluation data. 

\begin{figure*}[!t]
\centering
\fbox{\includegraphics[width=5in]{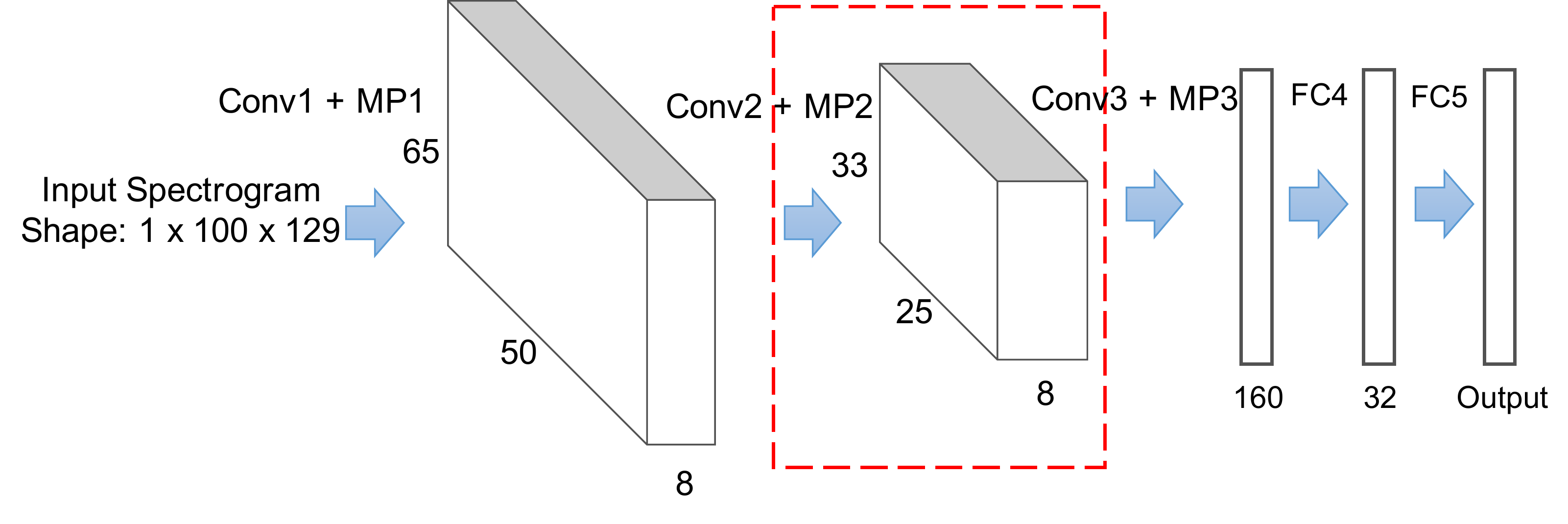}}
\caption{\it Architecture of the proposed model. The highlighted component shows a layer and its output feature map. For example, the shape of the feature map after the second convolutional and max pooling layer is $8 \times 25 \times 33$ (number of channels $\times$ time $\times$ frequency). Conv: Convolutional layer, FC: fully connected layer , MP: max pooling layer.}
\label{fig:fig2}
\end{figure*}

\section{Investigating CNN Architectures}

\subsection{Model 1 \cite{pindrop} } \label{model1_pindrop}

Now we replicate the CNN architecture of the second best performing deep-learning based system. Authors in \cite{pindrop} trained a CNN on hand-crafted features (CQCC+HFCC) to model the spoofing configuration in a multi-class setting. However, we train this CNN using two output targets to model the genuine and spoofed class distribution. 

Inspired from \cite{pindrop} we chose to use first one second of audio during training and evaluation. We use 512 point FFT, 512 window size and 10 ms hop size to produce a unified input spectrogram of $100 \times 257$ (time $\times$ frequency) dimension. Therefore, we train our CNN on these spectrograms. The details of the CNN architecture can be found in \cite{pindrop}. Since the implementation details of \cite{pindrop} is not disclosed in the paper, we chose to use parameter initialization and training approach described in section \ref{cnn_training_style0}. 

The performance of Model 1 is shown in Table \ref{all_results}. We only report the best system we found on this architecture, which show 5.47\% and 25.28\% EER on the development and evaluation data. However, this system uses a high dropout rate, with 90\% on the inputs of fully connected (FC) layer 1, 80\% on second FC layer, third FC layer inputs and 60\% on the inputs of the output layer.

\subsection{Model 2 \cite{thomasGrill} } \label{model2_bulbul}
This CNN architecture is motivated from the work of \cite{thomasGrill} on Birds Audio Detection (BAD) challenge 2017. Though the objective of the BAD and ASVspoof 2017 are completely different, they exhibit some similarity in the proposed test conditions: both focus on wild and diverse test conditions. Therefore, we adapt one of their CNN architecture, `Bulbul', to study if changing architecture helps improve the gap in generalization between the development and the evaluation data. 

The details of the `Bulbul' architecture can be found in \cite{thomasGrill}. We use split data, applying the algorithm described in appendix \ref{appendix:data_point}, during model training and testing. We use 256 point FFT, 1 seconds spectrogram window and shift window to obtain unified spectrogram of $100 \times 129$ dimension. At test time we take the average of the scores obtained for different spectrogram parts and compute the EER. We use the parameterization and training recipe as described in section \ref{cnn_training_style0}. 

We show the performance of Model 2 in Table \ref{all_results}. Though, we experimented with different dropouts, we found the best result using 50\% dropout on the fully connected (FC) layers inputs. This model give 4.52\% and 34.91\% EER on the development and the evaluation data respectively. However, the generalization gap between the two test sets is large.

\subsection{Model 3 } \label{model2_ours}

Our work so far have investigated different CNN architectures. These architectures ranges from medium to large in terms of the trainable parameters of the network. However, none of these systems showed similar-level of generalization on the development and evaluation data. Therefore, we now propose an architecture with smallest number of parameters that neither underfit nor overfit on the training data. This experiment seeks answer to our third question of section \ref{questions_for_study}.

This architecture has three convolutional layers and two fully connected layers. Each convolutional layer has 16 output filters (feature maps) and uses a small rectangular filter of $1\times 9$ with a stride of $1\times 1$ along time and frequency. We apply a max-pooling operation after each convolution layer. We use $3 \times 3$ kernel and $ 3 \times 3$ stride in all max-pooling layers. We use 32 neurons in the first fully connected layer with linear activation and two neurons in the output layer. All other layer use max-feature-map activation. We apply 50\% dropout on the fully connected layer inputs during training. We show the architecture of model 3 in Figure \ref{fig:fig2}. The input representation, model training and testing approach we used is similar as in Model 2 of section \ref{model2_bulbul}.

We show the performance of Model 3 in Table \ref{all_results}. Our proposed architecture seem to work quite well giving about 5\% EER on the development data. However, our model show a worse generalization on the evaluation data yielding an EER more than 30\%.

	

\section{Investigating Effect of Parameterization} \label{study3}

\subsection{Activation Function vs EER}

\begin{table}
\caption{\it EER \% for different activation function.}
\centering
\vspace{1.5 mm}
  \begin{tabular}{|c|c|c|}
  \hline 
  activation & dev & eval \\ [1ex]
  \hline \hline
  MFM & 4.98 & 33.11   \\
  RELU & 5.29 & 31.7   \\
  ELU  & 8.66 & 40.78   \\  
  \hline  
  \end{tabular}
  \label{activation_eer}
\end{table}

Our work so far have used MFM activation function inspired by the impressive results of \cite{lavrentyeva2017audio}. Here, we compare the performance of MFM with two other activations, RELU and ELU, that are often used in various deep learning tasks. We present the results in Table \ref{activation_eer}. ELU activation show the worse performance. On the development data, RELU and MFM seem to give similar performance. However, on the evaluation data RELU outperforms MFM and ELU. 

\subsection{Batch Size vs EER}
All our CNN experiments so far use 32 batch size. Here, we investigate how model performance compares when the network is trained using different batch sizes: 8, 16 and 64. We present the results in Table \ref{batch_size_results}. We see worse performance on both the development and evaluation data for 64 batch size. Similarly, batch size 16 does not seem to work well. Overall, we see an optimal performance for batch size of 32.

\begin{table}
\caption{\it EER\% for different batch sizes.}
\vspace{1.5mm}
\centering
  
\begin{tabular}{|c|c|c|}
  \hline
  batch size & dev & eval \\ [1ex]
  \hline \hline
   8   & 4.62 & 36.02   \\
  16  & 5.64 & 35.35   \\
  32  & 4.98 & 33.11   \\
  64  & 5.96 & 36.6   \\
  \hline  
  \end{tabular}
  \label{batch_size_results}
\end{table}

\subsection{Split Data vs EER} \label{split_experiments}

\begin{table}
\caption{\it EER \% using split spectrograms and single spectrogram.}
\vspace{1.5 mm}
\centering
  \begin{tabular}{|c|c|c|}
  \hline
  spectrogram & dev & eval \\ [1ex]
  \hline \hline
  split  &  4.98 & 33.11   \\
  single &  6.5  & 35.56     \\  
  \hline  
  \end{tabular}
  \label{split_data_results}
\end{table}

Our work explored utterance representation either by a single spectrogram or multiple splits using the approach described in appendix \ref{appendix:data_point}, for training the CNNs. Here, we compare the performance of these two representation. For single-spectrogram representation, we use three seconds audio (truncating/appending the original audio samples) to obtain $300 \times 129$ dimension spectrogram. Using split approach we obtain spectrograms of $100 \times 129$ dimension. For split, we chose one seconds spectrogram window and spectrogram shift. We used 256 point FFT in both cases. We show the results in Table \ref{split_data_results}. The model trained on split spectrograms outperforms single-spectrogram representation model on both the development and evaluation data.

\section{Summary and conclusion}

In this work, we discussed the ASVspoof challenge and its second edition (ASVspoof 2017) that focused on the replay attack detection. We described the best-performing systems from the challenge which were based on deep learning. We presented our motivation to implement an end-to-end model for the ASVspoof 2017 challenge and reported our experiments to implement one.  In our experiments,  we explored four end-to-end deep CNN architectures to design a replay attack detection system. We started with the state-of-the-art systems from the challenge, and later proposed a novel light-weight architecture.  In all our experiments the trained models failed to generalise in the evaluation dataset but achieved good performance in the development dataset. This intriguing result raises several interesting questions: why is it challenging to find an architecture that generalises in the evaluation dataset, how different are the data distributions between the subsets of the ASVspoof database.

Our current work does not use the newly patched ASVspoof 2017 database version 2. All our study is based on the version 1. We plan to use the new database for our future work that involves analysing the proposed model (Model 3) to understand what the model has learned about the genuine and spoofed signal. We aim to generate explanations for individual model predictions. Such an analysis will provide insight into the features maximally influencing a prediction.

\bibliographystyle{IEEEbib}
\bibliography{main}

\begin{thebibliography}{10}

\bibitem{ASV2017plan}
T.~Kinnunen et~al.,
\newblock ``{ASV}spoof 2017: Automatic speaker verification spoofing and
  countermeasures challenge evaluation plan.,'' 2017.

\bibitem{tomiSummaryPaper}
T.~Kinnunen et~al.,
\newblock ``The {ASV}spoof 2017 challenge: Assessing the limits of replay
  spoofing attack detection,''
\newblock {\em Proc. Interspeech 2017}, 2017.

\bibitem{LeCun2015}
Y.~LeCun, Y.~Bengio, and G.~Hinton,
\newblock ``{Deep Learning},''
\newblock {\em Nature}, vol. 521, no. 7553, pp. 436--444, 2015.

\bibitem{lavrentyeva2017audio}
G.~Lavrentyeva et~al.,
\newblock ``Audio replay attack detection with deep learning frameworks,''
\newblock {\em Proc. Interspeech 2017}, pp. 82--86, 2017.

\bibitem{pindrop}
Parav Nagarsheth, Elie Khoury, Kailash Patil, and Matt Garland,
\newblock ``Replay attack detection using {DNN} for channel discrimination,''
\newblock {\em Proc. Interspeech 2017}, pp. 97--101.

\bibitem{resnet}
Zhuxin Chen, Zhifeng Xie, Weibin Zhang, and Xiangmin Xu,
\newblock ``Resnet and model fusion for automatic spoofing detection,''
\newblock {\em Proc. Interspeech 2017}, pp. 102--106.

\bibitem{replay_using_highFreqFeats}
Marcin Witkowski, Stanislaw Kacprzak, Piotr Zelasko, Konrad Kowalczyk, and
  Jakub Galka,
\newblock ``Audio replay attack detection using high-frequency features,''
\newblock {\em Proc. Interspeech 2017}, 2017.

\bibitem{iiit_hyderabad}
K~N R~K Alluri, Sivanand Achanta, Sudarsana Kadiri, Suryakanth V.~Gangashetty,
  and Anil Vuppala,
\newblock ``{SFF} {A}nti-{S}poofer: {IIIT-H} {S}ubmission for {A}utomatic
  {S}peaker {V}erification {S}poofing and {C}ountermeasures {C}hallenge 2017,''
\newblock {\em Proc. Interspeech 2017}, pp. 107--111.

\bibitem{Montavon2018}
G.~Montavon, W.~Samek, and K.-R. M{\"u}ller,
\newblock ``{Methods for Interpreting and Understanding Deep Neural
  Networks},''
\newblock {\em Digital Signal Processing}, vol. 73, no. Supplement C, pp.
  1--15, 2018.

\bibitem{Simonyan2014}
K.~Simonyan, A.~Vedaldi, and A.~Zisserman,
\newblock ``{Deep Inside Convolutional Networks: Visualising Image
  Classification Models and Saliency Maps },''
\newblock in {\em Proc. ICLR}, 2014.

\bibitem{Zeiler2014}
M.~D {Zeiler} and R.~{Fergus},
\newblock ``{Visualizing and Understanding Convolutional Networks},''
\newblock in {\em Proc. ECCV}, 2014.

\bibitem{Mishra2017}
S.~Mishra, B.~L. Sturm, and S.~Dixon,
\newblock ``{Local Interpretable Model-Agnostic Explanations for Music Content
  Analysis},''
\newblock in {\em Proc. ISMIR}, 2017.

\bibitem{kinnunen2017reddots}
T.~Kinnunen et~al.,
\newblock ``Red{D}ots replayed: A new replay spoofing attack corpus for
  text-dependent speaker verification research,''
\newblock in {\em ICASSP 2017}. IEEE, 2017.

\bibitem{bhusanIcassp}
B.~Chettri and B.~L. Sturm,
\newblock ``A deeper look at gaussian mixture model based anti-spoofing
  systems,''
\newblock in {\em accepted in ICASSP 2018}. IEEE, 2018.

\bibitem{resnet_dataAugmentation}
Weicheng Cai, Cai Danwei, Wenbo Liu, Gang Li, and Ming Li,
\newblock ``Countermeasures for automatic speaker verification replay spoofing
  attack : On data augmentation, feature representation, classification and
  fusion,''
\newblock {\em Proc. Interspeech 2017}, pp. 17--21.

\bibitem{glorot}
X.~Glorot and Y.~Bengio,
\newblock ``Understanding the difﬁculty of training 
  networks,''
\newblock in {\em 13th International Conference on Artificial Intelligence and
  Statistics (AISTATS)}, 2010, vol.~9, pp. 249--256.

\bibitem{tensorflow2015-whitepaper}
M.~Abadi et~al.,
\newblock ``{TensorFlow}: Large-scale machine learning on heterogeneous
  systems,'' Software available from tensorflow.org.

\bibitem{bosaris2013brummer}
N.~Br{\"u}mmer and E.~D. Villiers,
\newblock ``The bosaris toolkit: Theory, algorithms and code for surviving the
  new dcf,''
\newblock {\em arXiv preprint arXiv:1304.2865}, 2013.

\bibitem{thomasGrill}
T.~Grill and J.~Schlüter,
\newblock ``Two convolutional neural networks for bird detection in audio
  signals,''
\newblock in {\em 2017 25th European Signal Processing Conference (EUSIPCO)},
  Aug 2017, pp. 1764--1768.

\end{thebibliography}

\begin{appendices}
\section{Data Split: Increasing the Data Points}
\label{appendix:data_point}
We propose a very simple technique that helps increase training data. We call this approach as `split data'. Using this approach we can generate large amount of training data points. Given an audio utterance $s$, we outline the algorithm for generating data points below. 

\begin{enumerate}
    \item Let $ l = length(s)$, be the original duration of $s$.
	\item Update $s$ by duplicating/truncating the samples such that $l_{new} = ceil(l)$.    
    \item Compute the log power magnitude spectrogram: $D = log |STFT(s)|^2$, where, D matrix has T number of frames and F frequency bins.
    \item Let $spec_{wind}$ and $wind_{shift}$ be the desired window and shift size (in time) respectively. Now, split $D$ into parts by moving $spec_{wind}$ by $wind_{shift}$.
    \item Return the list of spectrograms generated in step 4, where each spectrogram is of dimension $spec_{wind}$ x F.
    
\end{enumerate}

\end {appendices}

\end{document}